\documentclass[a4paper,12pt]{article}
\usepackage{jheppub}
\usepackage[utf8]{inputenc}
\usepackage{appendix}
\usepackage{mathtools}
\usepackage{amsmath}
\usepackage{comment}
\usepackage{amssymb}
\usepackage{braket}
\usepackage{hyperref}
\usepackage{url}
\usepackage{bibentry}
\numberwithin{equation}{section}

\def\raynote#1{{\color{red} #1}}
\newcommand\qvhard{Q_{V}^{\mathrm{hard}}}
\newcommand\qvsoft{Q_{V}^{\mathrm{soft}}}
\newcommand\qfhard{Q_{f}^{\mathrm{hard}}}
\newcommand\qfsoft{Q_{f}^{\mathrm{soft}}}

\newcommand\qf{Q_{f}}

\newcommand\qv{Q_{V}}

\newcommand{\braout}{\bra{\mathrm{out}}}
\newcommand{\ketin}{\ket{\mathrm{in}}}

\newcommand{\fpgrav}{a_{+}(E_{p}\hat{x})}

\newcommand{\vpgrav}{a_{+}(E_{q}\hat{y})}
\newcommand{\sa}{\braout \mathcal{S} \ketin}

\newcommand{\ep}{\epsilon_{p}}
\newcommand{\eq}{\epsilon_{q}}
\newcommand{\cd}{\cdot}
\newcommand{\sm}{\mathcal{S}}
\newcommand{\0}{(0)}
\newcommand{\1}{(1)}

\def\beq{\begin{equation}}
\def\eeq{\end{equation}}
\def\beqa{\begin{eqnarray}}
\def\eeqa{\end{eqnarray}}

\title{Double Soft Graviton Theorems and BMS Symmetries}

\author[a,b]{Anupam A.~H,}
\affiliation[a]{The Institute of Mathematical Sciences \\
	IV Cross Road, C.~I.~T.~Campus, \\
	Taramani, Chennai -- 600113, \\
	 Tamil Nadu, India \\}

\affiliation[b]{Homi Bhabha National Institute,\\
 Training School Complex,\\
Anushakti Nagar, \\
Mumbai -- 400094, \\
Maharashtra, India \\}

\emailAdd{anupam@imsc.res.in}

\author[a,b]{Arpan Kundu,}
\emailAdd{akundu@imsc.res.in} 

\author[c,d]{Krishnendu Ray}
\affiliation[c]{Chennai Mathematical Institute \\
H1, SIPCOT IT Park, \\
Siruseri, Kelambakkam, \\
Chennai -- 603103, \\
Tamil Nadu, India \\}

\affiliation[d]{International Centre for Theoretical Sciences \\
	Survey No. 151, Shivakote, \\
	Hesaraghatta, Hobli, \\
	Bengaluru North -- 560089 \\
	Karnataka, India \\}

\emailAdd{ray.krishnendu@cmi.ac.in}

\abstract{It is now well understood that Ward identities associated to the (extended) BMS algebra are equivalent to single soft graviton theorems. In this work, we show that if we consider nested Ward identities constructed out of two BMS charges, a class of double soft factorization theorems can be recovered. By making connections with earlier works in the literature, we argue that at the sub-leading order, these double soft graviton theorems are the so-called consecutive double soft graviton theorems. We also show how these nested Ward identities can be understood as Ward identities associated to BMS symmetries in scattering states defined around (non-Fock) vacua parametrized by supertranslations or superrotations.}

\begin{document}

\maketitle
\section{Introduction}\label{intro}

After the seminal work by Strominger \cite{strom}, there has been a flurry of activity towards understanding the role of a class of symmetries known as \emph{asymptotic symmetries} in gauge theories and gravity \cite{stromsublead,alok-1,alok-2, susy, qed, ym,alok-3,qed-hd,low, massless, alok:subleading, alok:another, alok:massive, alok:gauge-gravity, alok:new, st-high, vloop,thiago}. For theories containing massless particles of spin $1 \leq \, s \, \leq 2$, asymptotic symmetries are obtained by considering gauge transformations which do not fall off at infinity. Such large gauge transformations have non--trivial asymptotic charges and their conservation laws have non--trivial implications for the $\mathcal{S}$--matrix. 

For example, it has now become clear that the ``universal" soft theorems (i.e., those soft theorems whose structure is completely determined by gauge invariance \cite{broedel, bern}), such as the leading soft theorems in gauge theories and gravity, as well as the sub--leading soft theorem in gravity, are manifestations of Ward identities associated to a class of asymptotic symmetries (in 4 dimensions due to the infra--red divergences in these theories,  the cleanest statement can be made at the tree--level $\mathcal{S}$--matrix.). In the case of gravity, these symmetries are nothing but an infinite dimensional extension of the famous Bondi, Metzner, Sachs (BMS) group. 

However, factorization theorems in gauge theories and in quantum gravity have a richer structure. In the case of gravity, in a recent paper by Chakrabarti et al \cite{ashok}, it was shown that there exists a hierarchy of  factorization theorems when arbitrary but finite number of gravitons are taken to be soft in a scattering process. Of particular interest is the so called \emph{double soft graviton theorem}, which is a constraint on the scattering amplitude when two of the gravitons become soft. Such double soft theorems have a history in pion physics \cite{nima}. In the case of pions which are Goldstone modes of a spontaneously broken global non--abelian symmetry, double soft pion limits have an interesting structure. As was shown in \cite{nima}, if we consider a scattering amplitude in which two of the pions are taken to soft limit simultaneously, the scattering amplitude factorizes and the double soft theorem contains information about the structure of the (unbroken) symmetry generators. Due to the presence of an Adler zero, which ensures that single soft pion limit  vanishes, it is easy to see that there is no non--trivial factorization theorem if two pions are taken soft consecutively as opposed to when they are done so at the same rate. 

Double soft \emph{graviton} theorems are distinct in this regard. Not only is the simultaneous soft limit non--trivial and highly intricate, unlike the case of soft pions even the consecutive soft limit does not vanish and gives rise to factorization constraints on the scattering amplitude which are called the \emph{consecutive double soft theorems}.


In this paper we try to find an interpretation of such \emph{consecutive} double soft theorems as a consequence of Ward identities associated to the generalised BMS algebra\footnote{There are two known extensions of the BMS group in the literature. One is the ``extended BMS" \cite{stromsublead}, which is the semidirect product of supertranslations and the Virasoro group and the other is the ``generalised BMS" \cite{alok-2}, which is the semidirect product of supertranslations and Diff($S^{2}$). Each of them give rise to the same asymptotic charges and hence the same  Ward identities for the quantum gravity $\mathcal{S}$--matrix. Since, these Ward identities are the starting point of our analysis, this difference is irrelevant.}.\\

 The outline of this paper is as follows. In Section \ref{charges}, we recall the equivalence between leading and subleading soft graviton theorems and Ward identities associated to asymptotic symmetries \cite{CS,weinberg,strom,alok:subleading}. In Section \ref{Consecutive}, we explain  the consecutive double soft limit and how it gives rise to  a leading and two subleading consecutive double soft theorems.  In Section \ref{dst}, we propose asymptotic Ward identities, which, as we show in Appendix \ref{derieveconsecutive}, can be heuristically derived from Ward identities associated to Noether's charges \cite{avery}. In Appendix \ref{careful}, we discuss the conceptual subtleties associated to the domain of soft operators, which is an obstacle to the full rigorous derivation of one of the subleading consecutive double soft theorems from asymptotic symmetries. In Section \ref{scdst}, we present a formal derivation of this subleading consecutive double soft theorem from asymptotic symmetries.  We conclude with some remarks, which primarily focus on the key open question that pertaining to the study of the simultaneous double soft graviton theorem from the perspective of asymptotic symmetries. 

\section{Single Soft Graviton Theorems and Asymptotic Symmetries}\label{charges}
We begin by reviewing the derivations of the single soft graviton theorems (both leading and sub--leading) from asymptotic symmetries \cite{strom,alok:subleading}. In the process, we also define the notations that we use later. 

According to present understanding, the asymptotic symmetry group of gravity, acting on the asymptotic phase space of gravity is the ``Generalised BMS" group ---  it is a semidirect product of supertranslations and Diff($S^{2}$). They can be thought of as a local generalization of translations and the Lorentz group respectively. While the original BMS group \cite{bondi, sachs} is a semidirect product of supertranslations and $SL(2,\mathbb{C})$, in the generalised BMS group the $SL(2,\mathbb{C})$ symmetry is further extended to Diff($S^{2}$). Each of the supertranslations and Diff($S^{2}$) symmetry gives rise to conserved asymptotic charges, namely, the supertranslation charge ($Q_{f}$) and superrotation charge ($Q_{V}$) respectively. These charges are determined completely by the asymptotic ``free data" and are parametrized by an arbitrary function $f(z,\bar{z})$ and an arbitrary vector field  $V^{A}(z,\bar{z})$, respectively, both of which are defined on the conformal sphere at null infinity. By studying the algebra, one finds that supertranslations and superrotations form a closed algebra \cite{alok:new}. 

To define a symmetry of a gravitational scattering problem at the quantum level, these charges are elevated to a symmetry of the quantum gravity $\mathcal{S}$--matrix. Corresponding to each such symmetry one gets a Ward identity. In next two sections, we discuss that how the single soft graviton theorems are equivalent to Ward identities of generalised BMS charges.

\subsection{Leading Single Soft Graviton Theorem and Supertranslation Symmetry}
The leading single soft graviton theorem follows from the Ward identity of the supertranslation charge $Q_{f}$ \cite{strom}, which physically corresponds to the conservation of energy at each direction on the conformal sphere at null infinity.\footnote
{The seminal work \cite{strom} was based on the case when external states contain only massless particles. Generalisation to the case where external states can have massive particles was done in \cite{alok:massive}. } \\

The supertranslation charge $Q_{f}$ is given by \cite{strom}
\begin{equation}\label{classicalst}
	Q_{f}=\int \ du \ d^{2}z \ f \ \gamma_{z\bar{z}} \ N_{zz} \ N^{zz}+
	2\int du \ d^{2}z \ f \ \partial_{u}\Big(\partial_{z}U_{\bar{z}}+ \partial_{\bar{z}}U_{z}\Big)
\end{equation}
Here, $ \ U_{z}=-\frac{1}{2}D^{z}C_{zz} \ $,  and $ \ N_{zz}=\partial_{u}C_{zz} \ $ is the Bondi news tensor, where $C_{zz}$ is the ``free data''. The derivative $D^{z}$ is the covariant derivative \textit{w.r.t.} the 2--sphere metric. 

It is important to note that, the supertranslation charge $Q_{f}$ is characterized by the arbitrary function $f(z,\bar{z})$, where ($z$, $\bar{z}$) are coordinates on the conformal sphere at null infinity. Notice that, the first term in \eqref{classicalst} is quadratic in $C_{zz}$ while the second is linear in $C_{zz}$ --- these are conventionally referred as the ``hard part'' ($Q^{\mathrm{hard}}_{f}$) and the ``soft part'' ($Q^{\mathrm{soft}}_{f}$) of the supertranslation charge respectively.\\

In order to establish the equivalence between the supertranslation Ward identity and the leading single soft graviton theorem, the asymptotic charge \eqref{classicalst} is conjectured to be a symmetry of the quantum gravity $\mathcal{S}$--matrix \cite{strom}. As a result, one gets the Ward identity for supertranslation as:
\beq\label{super-duper}
\bra{\mathrm{out}}[Q_{f},\mathcal{S}]\ket{\mathrm{in}}=0 \Leftrightarrow \bra{\mathrm{out}}[Q^{\mathrm{soft}}_{f},\mathcal{S}]\ket{\mathrm{in}}=-\bra{\mathrm{out}}[Q^{\mathrm{hard}}_{f},\mathcal{S}]\ket{\mathrm{in}}
\eeq 
where in writing the above, the classical charges have been promoted to quantum operators. This quantization is carried out using the asymptotic quantization of $C_{zz}$ \cite{strom}, which expresses them in terms of graviton creation and annihilation operators. 

To evaluate \eqref{super-duper}, one needs to know the action of the hard and soft charges on the ``in" and ``out" states.
Let's begin by discussing the soft charge. Note that, we are working with Christodoulou--Klainerman (CK) spaces (which satisfy $D_{z}^{2}C_{\bar{z}\bar{z}}\vert_{\mathcal
{I^{+}}_{\pm}}=D_{\bar{z}}^{2}C_{zz}\vert_{\mathcal
{I^{+}}_{\pm}}$) \cite{strom}. This, together with the crossing symmetry of the scattering amplitude, allows one to write the  soft charge as:
\beqa\label{leading-soft}
Q^{\mathrm{soft}}_{f}=\lim_{E_{p} \rightarrow 0} \ \frac{E_{p}}{2\pi}\int d^{2}w \  D^{2}_{w} f(w,\bar{w}) \ a_{-}(E_{p},w,\bar{w}) \nonumber \\
= \lim_{E_{p} \rightarrow 0} \ \frac{E_{p}}{2\pi} \int d^{2}w \  D^{2}_{\bar{w}}f(w,\bar{w}) \ a_{+}(E_{p},w,\bar{w})
\eeqa
Hence, $Q^{\mathrm{soft}}_{f}\ket{\mathrm{in}}=0$. Here, $E_{p}$ is the energy of the soft graviton and $(w,\bar{w})$ characterizes its direction on the conformal sphere. 

The hard charge can also be evaluated in a similar procedure, finally giving the action on “in” and “out” states as:
\begin{equation}\label{HARD}
\begin{split}
Q_{f}^{\mathrm{hard}}\ket{\mathrm{in}} &=\sum\limits_{\mathrm{in}}E_{i} \ f(\hat{k_{i}})\ket{\mathrm{in}} 
\\ 
\bra{\mathrm{out}}Q_{f}^{\mathrm{hard}} &=\sum\limits_{\mathrm{out}}E_{i} \ f(\hat{k_{i}})\bra{\mathrm{out}}
\end{split}
\end{equation}
 Here, the sum $\sum\limits_{\mathrm{in}}$ and $\sum\limits_{\mathrm{out}}$ is over all the hard particles in the ``in" and ``out" states respectively, with energy $E_{i}=|\vec{k_{i}}|$ and the unit spatial vector $\hat{k_{i}}=\vec{k_{i}}/E_{i}$ characterizing the direction of $i^{\mathrm{th}}$ particle.

Using \eqref{HARD}, \eqref{leading-soft} and \eqref{super-duper} then, one obtains a factorization of the form:
\beqa\label{leading-ward-positive}
\lim_{E_{p} \rightarrow 0}\frac{E_{p}}{2\pi} \int d^{2}w \  D^{2}_{\bar{w}}f(w,\bar{w}) \ \bra{\mathrm{out}}a_{+}(E_{p},w,\bar{w}) \mathcal{S}\ket{\mathrm{in}}\nonumber\\
=-\bigg[\sum\limits_{\mathrm{out}}E_{i} \ f(\hat{k_{i}})-\sum\limits_{\mathrm{in}}E_{i} \ f(\hat{k_{i}})\bigg]&\sa
\eeqa\\

Structure of the terms in \eqref{leading-ward-positive} encourages one to ask whether this can be related to  Weinberg's soft graviton theorem \cite{weinberg}. This reads,
\beq\label{weinberg}
\lim_{E_{p} \rightarrow 0}E_{p} \bra{\mathrm{out} }a_{+} (E_{p},w,\bar{w}) \  \mathcal{S} \ket{\mathrm{in}}=\sum \limits_{i}\frac{(\epsilon^{+}(w,\bar{w}) \cdot k_{i})^{2}} {(p/E_{p})  \cdot k_{i}}\bra{\mathrm{out}}\mathcal{S}\ket{\mathrm{in}}
\eeq
where the soft graviton has energy $E_{p}$ and momentum $p$. Its direction is parametrized by ($w,\bar{w}$) and its polarization is given by $\epsilon^{+}(w,\bar{w})=1/\sqrt{2}(\bar{w},1,-i,-\bar{w})$. We adopt the notation:
\beq\label{lfactor}
\hat{S}^{(0)}(p;k_{i})\equiv \frac{1}{E_{k_{i}}} \frac{(\epsilon^{+}(w,\bar{w}) \cdot k_{i})^{2}} {(p/E_{p})  \cdot k_{i}}
\eeq 
with which, the leading soft factor in the r.h.s. of \eqref{weinberg} can be written as:
\beq\label{crazy-notation}
\sum \limits_{i}\frac{(\epsilon^{+}(w,\bar{w}) \cdot k_{i})^{2}} {(p/E_{p})  \cdot k_{i}}\equiv S^{(0)}(p;\{k_{i}\})\equiv \sum \limits_{i}S^{(0)}(p;k_{i}) \equiv \sum \limits_{i}E_{k_{i}} \ \hat{S}^{(0)}(p;k_{i})
\eeq
It is important to notice that the contribution to the soft factor $S^{(0)}(p;\{k_{i}\})$ from the $i^{\mathrm{th}}$ hard particle with momemtum $k_{i}$ and energy $E_{k_{i}}$, namely $S^{(0)}(p;k_{i})$, depends on the energy of the hard particle. But, $\hat{S}^{(0)}(p;k_{i})$ does not depend on $E_{k_{i}}$ ---  as written in \eqref{crazy-notation}, the energy dependence has been seperated out.

Now, consider  a hard particle of momentum $k$ parametrized by $(E,z,\bar{z})$. If one chooses
\begin{align}\label{leadfn}
		f(z,\bar{z}) = s(z,\bar{z};w,\bar{w}) \equiv \frac{1+w\bar{w}}{1+z\bar{z}} \cdot \frac{\bar{w}-\bar{z}}{w-z}
\end{align}
in \eqref{leading-ward-positive}, then the RHS of the soft theorem \eqref{weinberg} and the Ward identity \eqref{leading-ward-positive} match, since, 
\begin{align}
 \frac{(\epsilon^{+}(w,\bar{w}) \cdot k)^{2}} {(p/E_{p})  \cdot k}=-E_{k} \ s(z,\bar{z};w,\bar{w})
\end{align}
Further, the l.h.s. of the soft theorem \eqref{weinberg} and the Ward identity \eqref{leading-ward-positive} match because of the identity,
\begin{equation}\label{key}
	\begin{split}
		D^{2}_{\bar{z}}s(z,\bar{z};w,\bar{w})=2\pi\delta^{2}(w-z)
	\end{split}
\end{equation}
It is also possible to go from the soft theorem \eqref{weinberg} to the Ward identity \eqref{leading-ward-positive} by acting $(2\pi)^{-1}\int d^{2}w~ f(w,\bar{w}) D_{\bar{w}}^{2}$ on both sides of \eqref{weinberg}. In this case, the r.h.s. matches because of the identity:
\begin{equation}\label{key}
\begin{split}
D^{2}_{\bar{w}}s(z,\bar{z};w,\bar{w})=2\pi\delta^{2}(w-z)
\end{split}
\end{equation}
 Hence, the equivalence of the soft theorem and Ward identity is established. It
should also be noted that  Weinberg’s soft theorem for the negative helicity graviton is not an independent soft theorem and can be obtained through a similar derivation.

\subsection{Subleading Single Soft Graviton Theorem and Superrotation Symmetry}

The subleading single soft graviton theorem follows from the Ward identity of the superrotation charge $Q_{V}$ \cite{alok:subleading}, which physically corresponds to the conservation of angular momentum at each angle in a gravitational scattering process. This charge is given by:
\begin{align}\label{key1}\nonumber
		Q_{V}= \frac{1}{4} \int du \ d^{2}z \ \sqrt{\gamma} \ \partial_{u}C^{AB} \ \Big(\mathcal{L}_{V}&C_{AB}-\alpha~ C_{AB} + \alpha u \ \partial_{u}C_{AB}\Big) 
		\\
		+ \frac{1}{2} \int du \ &d^{2}z \ \sqrt{\gamma} \ \Big(C^{zz} \ D^{3}_{z}V^{z} + \ C^{\bar{z}\bar{z}} \ D^{3}_{\bar{z}}V^{\bar{z}} \Big)
\end{align}
where $\alpha=\frac{1}{2}\big(D_{z}V^{z}+D_{\bar{z}}V^{\bar{z}}\big)$ and $V^{A}(z,\bar{z})$ is an arbitrary vector field on the conformal sphere at null infinity. As usual, the covariant derivatives are {\it w.r.t.} the 2--sphere metric. As before, the first term is the “hard part” $Q^{\mathrm{hard}}_{V}$
and the second is the “soft part” $Q^{\mathrm{soft}}_{V}$ of the superrotation charge.\\

Proceeding in a manner similar to the case of supertranslation, the Ward identity for superrotations can be written as:
\beq\label{super-cool}
\bra{\mathrm{out}}[Q_{V},\mathcal{S}]\ket{\mathrm{in}}=0\Leftrightarrow \bra{\mathrm{out}}[Q^{\mathrm{soft}}_{V},\mathcal{S}]\ket{\mathrm{in}}=-\bra{\mathrm{out}}[Q^{\mathrm{hard}}_{V},\mathcal{S}]\ket{\mathrm{in}}
\eeq

Now, using the asymptotic quantization of the ``free data" and crossing symmetry one can write the soft superrotation charge as:
\beqa\nonumber
Q^{\mathrm{soft}}_{V}=
\frac{1}{4\pi i} \lim_{E_{p} \rightarrow 0} (1+E_{p} \ \partial_{E_{p}})&\\
\times\int d^{2}w \ \Bigg[V^{\bar{w}}   &\partial^{3}_{\bar{w}}  a_{+}(E_{p},w,\bar{w}) + V^{w} \ \partial^{3}_{w}  a_{-}(E_{p},w,\bar{w}) \Bigg]
\eeqa
Hence, $Q^{\mathrm{soft}}_{V}\ket{\mathrm{in}}=0$. Note that, unlike the previous case, due to the absence of a CK--like  condition, the action of $Q^{\mathrm{soft}}_{V}$ on the ``out" state gives gravitons of both helicities.
Also, the action of the hard superrotation charge gives: 
\begin{equation}\label{key}
\begin{split}
\bra{\mathrm{out}}Q^{\mathrm{hard}}_{V} &= i\sum\limits_{\mathrm{out}} J^{h_{i}}_{V_{i}}\bra{\mathrm{out}}
\\ 
Q^{\mathrm{hard}}_{V}\ket{\mathrm{in}} &= i\sum\limits_{\mathrm{in}} J^{-h_{i}}_{V_{i}}\ket{\mathrm{in}}
\end{split}
\end{equation}
Again, the sum $\sum\limits_{\mathrm{in}}$ and $\sum\limits_{\mathrm{out}}$ is over all the hard particles in the ``in" and ``out" states respectively, with the $i^{\mathrm{th}}$ particle having energy $E_{i}=|\vec{k_{i}}|$ and direction characterized by the vector $\hat{k_{i}}=\vec{k_{i}}/E_{i}$. A detailed expression of $J^{h_{i}}_{V_{i}}$ can be found in \cite{alok:subleading}.

As a result, one can write the Ward identity for superrotations \eqref{super-cool} as: 
\begin{equation}\label{superrotation-ward}
\begin{split}
&- \frac{1}{4\pi} \lim_{E_{p} \rightarrow 0} \ (1+E_{p} \ \partial_{E_{p}})
\\
&\quad \times \int d^{2}w \ \Bigg[V^{\bar{w}} \ \partial^{3}_{\bar{w}} \bra{\mathrm{out}} a_{+}(E_{p},w,\bar{w}) \ \mathcal{S} \ket{\mathrm{in}} + V^{w} \ \partial^{3}_{w} \bra{\mathrm{out}} a_{-}(E_{p},w,\bar{w}) \ \mathcal{S} \ket{\mathrm{in}} \Bigg]
\\
& \qquad \qquad \qquad \qquad \qquad \qquad \qquad \qquad \qquad \qquad = \Bigg[\sum_{\mathrm{out}} J^{h_{i}}_{V_{i}} - \sum_{\mathrm{in}} J^{-h_{i}}_{V_{i}}\Bigg] \bra{\mathrm{out}}\mathcal{S}\ket{\mathrm{in}} 
\end{split}
\end{equation}

Now, the Cachazo--Strominger (CS) subleading soft theorem reads \cite{CS}: 
\begin{equation}\label{cachazo-strominger}
\begin{split}
\lim_{E_{p}\rightarrow 0}(1+E_{p} \ \partial_{E_{p}})\bra{\mathrm{out}}&a_{+}(E_{p},w,\bar{w}) \ \mathcal{S}\ket{\mathrm{in}} 
\\
&=  \sum \limits_{i}\frac{\epsilon^{+}(w,\bar{w})\cdot k_{i}}{p\cdot k_{i}} \ \epsilon_{\mu}^{+}(w,\bar{w}) \ p_{\nu} \ J_{i}^{\mu\nu}\sa 
\end{split}
\end{equation}
 where, $J_{i}^{\mu\nu}$ is the angular momentum operator acting on the $i^{\mathrm{th}}$ hard particle. For further use, we adopt the notation:
 \beq\label{slfactor}
 S^{(1)}(p;{k_{i}})=\frac{\epsilon^{+}(w,\bar{w})\cdot k_{i}}{p\cdot k_{i}} \ \epsilon_{\mu}^{+}(w,\bar{w}) \ p_{\nu} \ J_{i}^{\mu\nu}
 \eeq
 Using this, the subleading soft factor in the r.h.s. of \eqref{cachazo-strominger} can be written as:
 \beq\label{crazy-notation-2}
  \sum \limits_{i}\frac{\epsilon^{+}(w,\bar{w})\cdot k_{i}}{p\cdot k_{i}} \ \epsilon_{\mu}^{+}(w,\bar{w}) \ p_{\nu} \ J_{i}^{\mu\nu}=\sum \limits_{i}S^{(1)}(p;{k_{i}}) = S^{(1)}(p;\{k_{i}\})
  \eeq 
Now, in the Ward identity \eqref{superrotation-ward}, if one chooses the vector field $V^{A}$ as:
\begin{align}\label{sleadfn}
V^{A}= K^{+}_{(w,\bar{w})} 
\equiv \frac{(\bar{z}-\bar{w})^{2}}{(z-w)} \ \partial_{\bar{z}}
\end{align}
the r.h.s. of the soft theorem \eqref{cachazo-strominger} and the Ward Idenity \eqref{superrotation-ward} match since:
\begin{align}\label{CSsoft}
\frac{\epsilon^{+}(w,\bar{w})\cdot k_{i}}{p \cdot k_{i}} \ \epsilon_{\mu}^{+}(w,\bar{w}) \ p_{\nu} \ J_{i}^{\mu\nu} = J^{i}_{K^{+}_{(w,\bar{w})}}
\end{align}
The l.h.s. of the soft theorem \eqref{cachazo-strominger} and the Ward identity \eqref{superrotation-ward} also match due to the identity:
\begin{align}\label{identity}
		\partial^{3}_{\bar{z}}\frac{(\bar{z}-\bar{w})^{2}}{(z-w)}=4\pi\delta^{2}(w-z)
\end{align}

To go from the CS soft theorem\eqref{cachazo-strominger} to the superrotation Ward identity \eqref{superrotation-ward} one acts the operator $-(4\pi)^{-1}\int d^{2}w  \ V^{\bar{w}}\partial_{\bar{w}}^{3}$ on both sides of \eqref{cachazo-strominger}. Then, using the linearity of $J_{V}$ in vector field $V$,
\begin{align}
-(4\pi)^{-1}\int d^{2}w \  V^{\bar{w}} \ \partial_{\bar{w}}^{3}J^{i}_{K^{+}_{(w,\bar{w})}}=-(4\pi)^{-1}J_{W}
\end{align}
and the identity,
\begin{align}
\partial^{3}_{\bar{w}}\frac{(\bar{z}-\bar{w})^{2}}{(z-w)}=-4\pi\delta^{2}(w-z)
\end{align}
one recovers Ward identity \eqref{superrotation-ward} with the vector field $V^{\bar{w}}\partial_{\bar{w}}$. The vector field $W$ in above expression is given by:
\begin{align}
W=\int V^{\bar{w}} \ \partial_{\bar{w}}^{3}K^{+}_{(w,\bar{w})}
\end{align}

Here, unlike the Ward identity for the leading case \eqref{leading-ward-positive}, it is important to note that the Ward identity for the subleading case \eqref{superrotation-ward}, contains both negative and positive helicity soft graviton amplitudes. To get a clear factorization, one of the components of vector field $V^{A}$ is chosen to be zero, depending upon which soft graviton helicity we want in the soft theorem. 


\section{Consecutive Double Soft Graviton Theorems (CDST)}\label{Consecutive}

Having reviewed the relationship between asymptotic symmetries and the single soft theorem, the next natural question is to ask if such a relationship holds between the generalised BMS algebra and double soft graviton theorems. These theorems (and its generalization to the multiple soft graviton case) have been studied previously using various methods including BCFW recursions \cite{nandan}, CHY amplitudes \cite{sen-chy, arnab1,arnab2,volovich} and Feynman diagram techniques \cite{ashok}. In a recent work \cite{lilin}, the authors have studied the symmetry foundations of the double soft theorems of certain classes of theories like the dilaton, DBI, and special Galileon. 

As has been analyzed in the literature, there are two kinds of double soft  graviton theorems depending upon the relative energy scale of the soft gravitons. The simultaneous soft limit is the one where soft limit is taken on both the gravitons at the same rate. It was shown in \cite{ashok}, that simultaneous soft limit  yields a universal factorization theorem. However, as we argue in Appendix \ref{derieveconsecutive}, from the perspective of Ward identities, it is the consecutive soft limits which arise rather naturally. Consecutive double soft graviton theorems (CDST) elucidate the factorization property of scattering amplitudes when the soft limit is taken on one of the gravitons at a faster rate than the other \cite{nandan}. We now review this factorization property when such soft limits are taken  and show that they give rise to three CDSTs. The first one, we refer to as the leading CDST which is the case where the leading soft limit is taken on both the soft gravitons. The remaining two theorems refer to the case where the leading soft limit is taken with respect to one of the gravitons and the subleading soft limit is taken with respect to the other.

We begin with a $(n+2)$ particle scattering amplitude denoted by $\mathcal{A}_{n+2}(q,p,\{k_{m}\})$ where $p~$,$~q$ are the momenta of the two gravitons which will be taken to be soft and $\{k_{m}\}$ is the set of momenta of the $n$ hard particles. Consider the consecutive limit where the soft limit is first taken on graviton with momentum $q$, keeping all the other particles momenta unchanged and then a soft limit is taken on the graviton with momentum $p$.

 
Using the single soft factorization, the scattering amplitude $\mathcal{A}_{n+2}(q,p,\{k_{m}\})$ can be written as:
\begin{align}\label{cons2}\nonumber
\mathcal{A}_{n+2}(q,p,\{k_{m}\})=\Bigg[\sum\limits_{i}\frac{E_{k_{i}}}{E_{q}}& \ \hat{S}^{(0)}(q;k_{i})+\frac{E_{p}}{E_{q}} \ \hat{S}^{(0)}(q;p)\\
+\sum\limits_{i}&S^{(1)}(q;k_{i})+S^{(1)}(q;p)\Bigg]\mathcal{A}_{n+1}(p,\{k_{m}\})+\mathcal{O}(E_{q})
\end{align}
where $\mathcal{A}_{n+1}(p,\{k_{m}\})$ is the $n+1$ particle scattering amplitude. It is important to recall the notations used here, which we explained in Section \ref{charges} (\eqref{crazy-notation}, \eqref{slfactor}). As mentioned, $S^{(1)}(q;k_{i})$ is the contribution to the subleading soft factor with soft momentum $q$ with $k_{i}$ being the $i^{\mathrm{th}}$ hard particle. Similarly $\hat{S}^{(0)}(q;k_{i})$ denotes the contribution to the subleading soft factor with soft momentum $q$ with $k_{i}$ being the $i^{\mathrm{th}}$ hard particle, with energy dependences \textit{w.r.t.} both the soft and hard particles seperated out. $\hat{S}^{(0)}(q;p)$ and $S^{(1)}(q;p)$ denote similar contributions to the soft factor where the graviton with momentum $p$ is treated as hard {\it w.r.t.} the graviton with momentum $q$.


Now,  the amplitude $\mathcal{A}_{n+1}(p,\{k_{m}\})$ further factorizes as:
  \begin{align}\label{cons1}
\mathcal{A}_{n+1}(p,\{k_{m}\})=\Bigg[\sum\limits_{i}\frac{E_{k_{i}}}{E_{p}} \ \hat{S}^{(0)}(p;k_{i})+\sum\limits_{i}S^{(1)}(p;k_{i})\Bigg]\mathcal{A}_{n}(\{k_{m}\})+\mathcal{O}(E_{p})
\end{align}
Note that, according to our notation, $S^{(1)}(p;k_{i})$ is the contribution to the subleading soft factor with soft momentum $p$ and $k_{i}$ is the $i^{\mathrm{th}}$ hard particle. Again, $\hat{S}^{(0)}(p;k_{i})$ denotes the contribution to the subleading soft factor with soft momentum $p$ and $k_{i}$ the  $i^{\mathrm{th}}$ hard particle, with energy dependences \textit{w.r.t.} both the soft and the hard particles seperated out.

Substituting \eqref{cons1} in \eqref{cons2}, we get the factorization of the $(n+2)$ particle amplitude containing two soft gravitons in terms of the amplitude of the $n$ hard particles (up to subleading order in energy of the individual soft particles). 
\begin{align}\nonumber
\mathcal{A}_{n+2}&(q,p,\{k_{m}\})=\nonumber
\\&\nonumber\bigg[\frac{1}{E_{p}E_{q}}\sum\limits_{i,j}E_{k_{i}} \ E_{k_{j}} \  \hat{S}^{(0)}(q;k_{i}) \ \hat{S}^{(0)}(p;k_{j})
+\sum\limits_{i,j}\frac{E_{k_{i}}}{E_{q}} \ \hat{S}^{(0)}(q;k_{i}) \ S^{(1)}(p;k_{j})
\\&~~~~~~~~~~\nonumber + \sum\limits_{i}\frac{E_{k_{i}}}{E_{q}} \ \hat{S}^{(0)}(q;p) \ \hat{S}^{(0)}(p;k_{i})~+
\sum\limits_{i,j}S^{(1)}(q;k_{i})  \ \frac{E_{k_{j}}}{E_{p}} \ \hat{S}^{(0)}(p;k_{j}) 
\\&~~~~~~~~~~~~~~+
S^{(1)}(q;p)\sum\limits_{i}\frac{E_{k_{i}}}{E_{p}} \ \hat{S}^{(0)}(p;k_{i})\bigg] \mathcal{A}_{n}(\{k_{m}\}) +\mathcal{O}(E_{p})+\mathcal{O}(E_{q})
\end{align}
This expansion contains three types of terms. The first type scales as $1/(E_{p}E_{q})$ (and hence gives rise to a pole in both the soft graviton energies), giving the leading contribution to the factorization. The second and the third type of terms scale as $E_{q}^{0}/E_{p}$ and $E_{p}^{0}/E_{q}$ respectively, both contributing to the subleading order of the factorization.

The leading order contribution, described above, is:
\begin{align}
\bigg[\frac{1}{E_{p}E_{q}}\sum\limits_{i,j}E_{k_{i}} \ E_{k_{j}} \ \hat{S}^{(0)}(q;k_{i}) \ \hat{S}^{(0)}(p;k_{j})\bigg]~\mathcal{A}_{n}(\{k_{m}\})
\end{align}
This gives the \emph{leading} CDST as:
\begin{align}\label{leadCDSL}
\lim_{E_{p}\rightarrow 0}E_{p}\lim_{E_{q}\rightarrow 0}E_{q}~\mathcal{A}_{n+2}(q,p,\{k_{m}\})=\bigg[~S^{(0)}(q;\{k_{i}\})  \ S^{(0)}(p;\{k_{j}\})\bigg]~\mathcal{A}_{n}(\{k_{m}\})
\end{align}

As is evident, the leading double soft factor is just the product of the individual leading soft factors. One obtains this same theorem in the case of the simultaneous double soft limit as well \cite{ashok,nandan,arnab1,arnab2}.  In Section \ref{lcsl}, we show that this soft theorem matches with the result derived from the Ward identity of two supertranslation charges \eqref{dlst}.

Let us now consider the subleading soft limit. At this order of factorization we have four terms:
\begin{align}\label{subCDSL}\nonumber
\bigg[\sum\limits_{i,j}& \ \frac{E_{k_{i}}}{E_{q}} \ \hat{S}^{(0)}(q;k_{i}) \ S^{(1)}(p;k_{j}) + \sum\limits_{i}\frac{E_{k_{i}}}{E_{q}} \ \hat{S}^{(0)}(q;p) \ \hat{S}^{(0)}(p;k_{i}) \\~+
&\sum\limits_{i}S^{(1)}(q;k_{i}) \sum\limits_{j}\frac{E_{k_{j}}}{E_{p}} \ \hat{S}^{(0)}(p;k_{j}) +
S^{(1)}(q;p)\sum\limits_{i}\frac{E_{k_{i}}}{E_{p}} \ \hat{S}^{(0)}(p;k_{i})\bigg] \mathcal{A}_{n}(\{k_{m}\})
\end{align}
Notice that the first two terms in (\ref{subCDSL}) scale with soft graviton energies as  $E_{p}^{0}/E_{q}$ and the second two terms scale as $E_{q}^{0}/E_{p}$.

From the first two terms of (\ref{subCDSL}), one gets a subleading CDST.
\begin{align}\nonumber\label{proj1}
\lim_{E_{p}\rightarrow 0}(1+E_{p}\partial_{E_{p}})\lim_{E_{q}\rightarrow 0}E_{q}& \ \mathcal{A}_{n+2}(q,p,\{k_{m}\})\\=\bigg[& S^{(0)}(q;\{k_{i}\}) \ S^{(1)}(p;\{k_{j}\}) + \mathcal{N}(q;p;\{k_{i}\})\bigg] \mathcal{A}_{n}(\{k_{m}\})
\end{align}

Here, the first term is the product of single soft factors \eqref{crazy-notation}, \eqref{crazy-notation-2}, appearing in the leading and subleading single soft theorems respectively. The second term in the r.h.s of \eqref{proj1}  contains a single sum over the set of hard particles as opposed to the first term which is the product of single soft factors and contains two sums over the set of hard particles. Such terms are usually referred to as ``contact terms" in the literature. One can evaluate this contact term as:
\begin{align}
\mathcal{N}(q;p;\{k_{i}\}) =~\hat{S}^{(0)}(q;p) \  S^{(0)}(p;\{k_{i}\})
=\sum\limits_{i}  \frac{(\eq \cdot \tilde{p})^{2}} {\tilde{q}  \cdot \tilde{p}} \cdot  \frac{(\ep \cdot k_{i})^{2}} {\tilde{p} \cdot k_{i}}
\end{align}
where $\tilde{p}=p/E_{p}=(1,\hat{p})$ and similarly, $\tilde{q}= q/E_{q}=(1,\hat{q})$.  $\ep$ and $\eq$ refer to  the polarisations of soft gravitons with momentum $p$ and $q$ respectively. This is the well known consecutive double soft graviton theorem \cite{nandan} . 

\subsection*{A Different Consecutive Limit.}
We now take a different limit in eq.(\ref{subCDSL}) and show how it leads to a distinct factorization theorem. 
From the last two terms in (\ref{subCDSL}) one gets:
\begin{align}\label{contact}
&\lim_{E_{p}\rightarrow 0} E_{p}\lim_{E_{q}\rightarrow 0}(1+E_{q} \ \partial_{E_{q}}) \ \mathcal{A}_{n+2}(q,p,\{k_{m}\})\\\nonumber
&=
\bigg[\sum\limits_{i}S^{(1)}(q;k_{i})\sum\limits_{j}E_{k_{j}} \ \hat{S}^{(0)}(p;k_{j}) +
\lim_{E_{p}\rightarrow 0}E_{p} \ S^{(1)}(q;p)\sum\limits_{i}\frac{E_{k_{i}}}{E_{p}} \ \hat{S}^{(0)}(p;k_{i})\bigg] \mathcal{A}_{n}(\{k_{m}\})
\end{align}
Now, $S^{(1)}(q;k_{i})$ contains the angular momentum operator of the $i^{\mathrm{th}}$ hard particle, and thus acts on $E_{k_{j}}\hat{S}^{(0)}(p;{k_{j}})$, as well as the $n$ particle amplitude $\mathcal{A}_{n}(\{k_{m}\})$. However, $S^{(1)}(q;p)$ does not depend on the set of hard particles labelled by momentum $\{k_{m}\}$. Hence $S^{(1)}(q;p)$ acts only on the soft factor, and one can finally write the subleading CDST as:
\begin{align}\nonumber\label{contact2}
&\lim_{E_{p}\rightarrow 0 }E_{p} \lim_{E_{q}\rightarrow 0}(1+E_{q} \ \partial_ {E_{q}}) \ \mathcal{A}_{n+2}(q,p,\{k_{m}\})\\
=&\Bigg[S^{\0}(p;\{k_{i}\}) \ S^{\1}(q;\{k_{j}\}) + \mathcal{M}_{1}(q;p;\{k_{i}\}) + \mathcal{M}_{2}(q;p;\{k_{i}\})\Bigg]\mathcal{A}_{n}(\{k_{m}\})
\end{align}
Similar to the other subleading CDST \eqref{proj1}, the first term in the r.h.s. of \eqref{contact2} is product of single soft factors. However, the important difference is that the role of the soft gravitons with momentum $p$ and $q$ is interchanged in the first term of \eqref{contact2} and the first term of \eqref{proj1}. Here, $\mathcal{M}_{1}(q;p;\{k_{i}\})$ and $\mathcal{M}_{2}(q;p;\{k_{i}\})$ are contact terms which can be expressed as follows:
\begin{align}\label{ttt-1}
\mathcal{M}_{1}(q;p;\{k_{i}\})=\sum\limits_{i}& S^{(1)}(q;k_{i})\Big(E_{k_{i}} \ \hat{S}^{\0}(p;k_{i})\Big)=\sum\limits_{i} S^{(1)}(q;k_{i})\Big(S^{\0}(p;k_{i})\Big)\nonumber\\
=\sum\limits_{i}&\bigg[-\frac{(\eq \cd k_{i})^{2}(\ep\cd k_{i})^{2}(p\cd q)}{(q\cd k_{i})(p\cd k_{i})^{2}}+\frac{(\eq \cd k_{i})(\eq \cd p)(\ep\cd k_{i})^{2}}{(p\cd k_{i})^{2}}\nonumber\\
 &+2\frac{(\eq\cd k_{i})^{2}(\ep\cd k_{i})(\ep \cd q)}{(p\cd k_{i})(q\cd k_{i})}-2\frac{(\eq\cd k_{i})(\ep\cd\eq)(\ep\cd k_{i})}{(p\cd k_{i})}\bigg]
 \end{align}
and,
\begin{align}\label{ttt-2}
\mathcal{M}_{2}(q;p;\{k_{i}\})&=\sum\limits_{i}\lim\limits_{E_{p}\rightarrow 0} E_{p}~ S^{(1)}(q;p)\Bigg(\frac{E_{k_{i}}}{E_{p}} \ \hat{S}^{\0}(p;k_{i}) \Bigg)\nonumber\\
=\sum\limits_{i}&\bigg[\frac{(\eq\cd \tilde{p})(\eq\cd k_{i})(\ep\cd k_{i})^{2}}{(\tilde{p}\cd k_{i})^{2}}
-\frac{(\eq\cd \tilde{p})^{2}(\ep\cd k_{i})^{2}(q\cd k_{i})}{(\tilde{p}\cd k_{i})^{2}(\tilde{p}.q)}\nonumber\\
-&  \ 2 \ \frac{(\eq\cd \tilde{p})(\eq\cd k_{i})(\ep \cd q)(\ep \cd k_{i})}{(\tilde{p}.q)(\tilde{p}.k_{i})}+ 2 \ \frac{(\eq \cd \tilde{p})(\eq\cd\ep)(\ep \cd k_{i})(q.k_{i})}{(\tilde{p}.q)(\tilde{p}.k_{i})}\bigg]
\end{align}
Again, $\tilde{p}=p/E_{p}=(1,\hat{p})$ and $\ep$ and $\eq$ refer to the polarisation of soft gravitons with momentum $p$ and $q$ respectively.


In \cite{nandan}, the authors have considered similar consecutive limits for the double soft graviton and gluon amplitudes. There, they have imposed a gauge condition $\epsilon_{p}\cdot q=0$ and	$\epsilon_{q}\cdot p=0$. However, our analysis proceeded without imposing any particular gauge condition. With the specific gauge condition used in \cite{nandan}, a few of the terms like $\hat{S}^{(0)}(q;p)$ and $S^{(1)}(q;p)$ drop out from the CDST result that we have obtained at the subleading level and we recover their result. This serves as a consistency check for our calculation.

 One can also verify the consistency of both the consecutive limits with the general result which was given in \cite{ashok}. That is, both the CDST \eqref{proj1} and \eqref{contact2} are special cases of the double soft limit  in \cite{ashok}. The CDST \eqref{proj1} can be recovered by imposing the condition $E_{p}\gg E_{q}$ on the result of \cite{ashok} and  taking the leading limit in $E_{q}$ and subleading limit in $E_{p}$. Similarly, the CDST \eqref{contact2} can be obtained by imposing the  the same $E_{p}\gg E_{q}$ condition, but taking the leading limit in $E_{p}$ and subleading limit in $E_{q}$.

In the subsequent sections, we will argue that these soft theorems are equivalent to Ward identities of asymptotic symmetries when the scattering states are defined with respect to super--translated or super--rotated vacua.

\section{ CDST and Asymptotic Symmetries}
\subsection{Introduction}\label{dst}
Having reviewed the relationship between Ward identities associated to the asymptotic symmetries and single soft graviton theorems, we now ask if  there are Ward identities in the theory which are equivalent to the double soft graviton theorems at the leading and sub--leading order.  In particular, we look for Ward identities that will lead us to the consecutive double soft theorems (CDST).  Let us consider the family of Ward identities whose general structure is:
\begin{align}
\bra{\mathrm{out}} \big[  Q_{1},  [Q_{2},\mathcal{S}] \big]\ket{\mathrm{in}}=0
\end{align}
where both $Q_{1}$ and $Q_{2}$ are either both supertranslation charges or $Q_{1}$ is a supertranslation charge and 
$Q_{2}$ is a superrotation charge.\footnote{The alternate case where $Q_{1}$ is superrotation charge and $Q_{2}$ is supertranslation charge is riddled with conceptual subtleties which remain unresolved ---  we return to this in Appendix \ref{careful}.}

Following \cite{avery}, we present a  derivation of this proposed Ward identity in Appendix \ref{derieveconsecutive}. In the following sections, we show that such a proposal leads to the consecutive double
soft theorems discussed in Section \ref{Consecutive}. Depending on the choice of charges one gets the leading as well as the subleading consecutive double soft theorems.

\subsection{Leading CDST and Asymptotic Symmetries }\label{lcsl}
\subsubsection{Ward Identity from Asymptotic Symmetries}
Following the discussion in Section \ref{dst}, we explore the factorization arising from two supertranslation charges, $Q_{f}$ and $Q_{g}$ characterized by arbitrary functions $f (z,\bar{z})$ and $g(z, \bar{z})$, on the conformal sphere.
We start with:
\begin{align}\label{leading-cdsl}
\bra{\mathrm{out}} \big[  Q_{f},  [Q_{g},\mathcal{S}] \big]\ket{\mathrm{in}}=0
\end{align}
 Proceeding in a manner similar to the single soft case in Section \ref{charges}, we can write $Q_{f}$ and $Q_{g}$ as sum of hard and soft charges as:
\begin{align}\label{key}
Q_{f}=Q^{\mathrm{hard}}_{f}+Q^{\mathrm{soft}}_{f} \ \ ,\ \ \  Q_{g}=Q^{\mathrm{hard}}_{g}+Q^{\mathrm{soft}}_{g}
\end{align}
Thus, the Ward identity \eqref{leading-cdsl} becomes:
\begin{equation}\label{lds2}
\begin{split}
\bra{\mathrm{\mathrm{out}}} \big[Q^{\mathrm{hard}}_{f}, [Q^{\mathrm{hard}}_{g}& ,\mathcal{S}] \big] \ket{\mathrm{in}}  + \bra{\mathrm{out}} \big[Q^{\mathrm{hard}}_{f} ,[Q^{\mathrm{soft}}_{g}, \mathcal{S}]\big] \ket{\mathrm{in}} 
\\
&+ \bra{\mathrm{out}} \big[Q^{\mathrm{soft}}_{f}, [Q^{\mathrm{hard}}_{g}, \mathcal{S}]\big] \ket{\mathrm{in}} + \bra{\mathrm{out}} \big[Q^{\mathrm{soft}}_{f}, [Q^{\mathrm{soft}}_{g}, \mathcal{S}]\big] \ket{\mathrm{in}} = 0
\end{split}
\end{equation}

Now using the Ward identity of supertranslation, namely $[Q_{g}^{\textrm{soft}},S]=-[Q_{g}^{\textrm{hard}},S]$, the first and the second terms cancel each other. One may be tempted to cancel the third and fourth terms, on similar lines. However, we contend that this isn't quite correct as the action of $\qfsoft$ maps ordinary the Fock vaccuum to a supertranslated vaccuum state parametrised by $f$. As a result, we are really looking at the following Ward identity. 
\begin{equation}
\bra{\mathrm{out}, f} [Q_{g},S] \ket{\mathrm{in}} =\ 0
\end{equation}
where $\ket{\mathrm{out}, f}$ is a finite energy state defined with respect to the super--translated vacuum. The ``in" state is defined w.r.t standard Fock Vacuum because of our prescription $Q_{f}^{\textrm{soft}}\vert\textrm{in}\rangle\ =\ 0$.
We can re--write the above identity as:
\begin{align}
\braout \big[\qfsoft,[Q_{g}^{\textrm{soft}},\mathcal{S}] \big]\ketin= - \braout \big[\qfsoft,[Q_{g}^{\textrm{hard}},\mathcal{S}] \big] \ketin
\end{align}
Now using the Jacobi identity among $\qfsoft$, $Q_{g}^{\textrm{hard}}$ and $\sm$, the commutation relation $[\qfsoft,Q_{g}^{\textrm{hard}}]=0$, and the single soft Ward identity, we can finally write 
\begin{align}\label{lds3}
\braout \big[\qfsoft,[Q_{g}^{\textrm{soft}},\mathcal{S}] \big]\ketin=\braout \big[Q_{g}^{\textrm{hard}},[\qfhard,\mathcal{S}] \big]\ketin
\end{align}.

Using the (known) action of charges on external states in \eqref{lds3}
 we finally arrive at the Ward identity:
\begin{align}\label{final1}\nonumber
\lim_{E_{p} \rightarrow0} \ \frac{E_{p}}{2\pi} \ \lim_{ E_{q}\rightarrow0} \ \frac{E_{q}}{2\pi} \int d^{2}w_{1} \ d^{2}w_{2} ~& D_{\bar{w}_{1}}^{2}f(w_{1},\bar{w_{1}}) \    D_{\bar{w}_{2}}^{2}g(w_{2},\bar{w_{2}})  \\\nonumber& \times\bra{\mathrm{out}} a_{+}(E_{p},w_{1},\bar{w}_{1}) \  a_{+}(E_{q},w_{2},\bar{w}_{2}) \ \mathcal{S}\ket{\mathrm{in}}
\\
 = \Bigg[\sum_{\mathrm{out}} f(\hat{k_{i}})E_{i} - \sum_{\mathrm{in}}  f(\hat{k_{i}}) E_{i}\Bigg] &\Bigg[\sum_{\mathrm{out}} g(\hat{k_{j}})E_{j} - \sum_{\mathrm{in}} g(\hat{k_{j}})E_{j} \Bigg] \bra{\mathrm{out}} \mathcal{S} \ket{\mathrm{in}}
\end{align}
 The factorization above is just the product of two factors of the type obtained from the Ward identity for supertranslation \eqref{leading-ward-positive}. It is natural therefore to expect that the soft theorem we obtain from \eqref{final1} will also be the product of two leading single soft factors. In the next section, we show that this is indeed true.   

\bigskip
\subsubsection{From Ward Identity To Soft Theorem}

From the factorization obtained in \eqref{final1} from the Ward identity with two supertranslation charges, we try to understand what soft theorem follows from it. Motivated from the single soft case, we make the choices for arbitrary function $f$ and $g$ on the conformal sphere as:
\begin{align}\label{key}
f(w_{1},\bar{w_{1}}) = s(w_{1},\bar{w_{1}};w_{p},\bar{w_{p}}) \ \ , \ \ \ g(w_{2},\bar{w_{2}}) = s(w_{2},\bar{w_{2}};w_{q},\bar{w_{q}})
\end{align}
where the definition of the functions $s(w_{1},\bar{w_{1}};w_{p},\bar{w_{p}})$ and $s(w_{2},\bar{w_{2}};w_{q},\bar{w_{q}})$ can be read from \eqref{leadfn}. Substituting these choices in \eqref{final1}, we finally get:
\begin{align}\label{dlst}\nonumber
\lim_{E_{p} \rightarrow 0} E_{p} \lim_{E_{q}\rightarrow 0} E_{q} \ \bra{\mathrm{out}} a_{+}(E_{p},&w_{p},\bar{w_{p}}) \ a_{+}(E_{q},w_{q},\bar{w_{q}}) \ \mathcal{S} \ket{\mathrm{in}} \\
&= \bigg[S^{(0)}(q;\{k_{i}\}) \ S^{(0)}(p;\{k_{j}\})\bigg]\sa
\end{align}
This is the same as the leading double soft theorem \eqref{leadCDSL} for the case of two positive helicity soft gravitons with momenta $p$ and $q$, localized at $(w_{p},\bar{w_{p}})$ and $(w_{q},\bar{w_{q}})$ respectively, on the conformal sphere. Although we have chosen both the soft graviton helicities to be positive in the above, one can do a similar analysis for both the helicities being negative or one positive and one negative, and a similar result holds. This provides the equivalence of the leading CDST and the Ward identity \eqref{leading-cdsl}.

We have thus shown that the leading order double soft graviton theorem is equivalent to the supertranslation Ward identity when this identity is evaluated in a Hilbert space built out of a super--translated vacuum that containing a single soft graviton.


\subsection{Subleading CDST and Asymptotic Symmetries}\label{scsl}
\subsubsection{Ward Identity from Asymptotic Symmetries}\label{scsl2a}
As motivated in Section \ref{dst}, and derived in Appendix \ref{derieveconsecutive}, we now analyze with the Ward identity corresponding to one supertranslation charge (characterized by arbitrary function $f$)  and one superrotation charge (characterized by vector field $V^{A}$): 
\begin{align}\label{right-sub-ward}
\braout \big[\qf,[\qv,\sm] \big]\ketin=0
\end{align}
We begin by writing the charges as sum of hard and soft charges: 
\begin{equation}\label{slds2}
\begin{split}
\bra{\mathrm{out}} \big[Q^{\mathrm{hard}}_{f}, [Q^{\mathrm{hard}}_{V}& ,\mathcal{S}]\big] \ket{\mathrm{in}}  + \bra{\mathrm{out}} \big[Q^{\mathrm{hard}}_{f} ,[Q^{\mathrm{soft}}_{V}, \mathcal{S}]\big] \ket{\mathrm{in}} 
\\
+ \bra{\mathrm{out}}& \big[Q^{\mathrm{soft}}_{f}, [Q^{\mathrm{hard}}_{V}, \mathcal{S}]\big] \ket{\mathrm{in}} + \bra{\mathrm{out}} \big[Q^{\mathrm{soft}}_{f}, [Q^{\mathrm{soft}}_{V}, \mathcal{S}]\big] \ket{\mathrm{in}} = 0
\end{split}
\end{equation}

Now, using the Ward identity for superrotation, namely $[\qvsoft,\sm]=-[\qvhard,\sm]$, the first and the second term of \eqref{slds2} cancel each other. Again, one may be tempted to cancel the third and the fourth term  of \eqref{slds2} instead, using the same superrotation Ward identity. However if we do not cancel them, we are led to
\begin{equation}\label{key}
\begin{split}
\bra{\mathrm{out}} Q_{f}^{\textrm{soft}} \ [Q_{V},S] \ket{\mathrm{in}} = 0
\\
\bra{\mathrm{out}, f}[Q_{V},S] \ket{\mathrm{in}}= 0
\end{split}
\end{equation}
Whence not cancelling the third and forth terms in (\ref{slds2}) is tantamount to considering superrotation Ward identity in scattering states which are exitations around supertranslated vacuua. As we show below, it is precisely the Ward identity \\
$\bra{\mathrm{out}, f} [Q_{V},\mathcal{S}] \ket{\mathrm{in}} = 0$ that leads to a specific double soft graviton theorem.

Hence the above identity \eqref{slds2} reduces to, 
\begin{equation}\label{ward3al}
\begin{split}
\braout \big[\qfsoft,[\qvsoft, &\sm] \big]\ketin 
\\
&=-\braout \big[\qfsoft,[\qvhard,\sm] \big]\ketin
\\
&=-\braout\qfsoft \ \qvhard \ \sm\ketin+\braout\qfsoft \ \sm \ \qvhard\ketin
\end{split}
\end{equation}
Using the known action of the soft and hard charges, first term in the r.h.s. of \eqref{ward3al} can be written as:
\begin{align}\nonumber
\braout\qfsoft \ &\qvhard \ \sm \ketin
\\  \nonumber
&= \frac{1}{2\pi }\lim_{E_{p}\rightarrow 0}\int d^{2}{w_{1}}~D^{2}_{\bar{w_{1}}}f~E_{p}\braout a_{+}(E_{p}\hat{x}) \ \qvhard \ \sm\ketin \\
& =\frac{i}{2\pi}\lim_{E_{p}\rightarrow 0}\int d^{2}{w_{1}}~D^{2}_{\bar{w_{1}}}f~E_{p}\Big(\sum_{\mathrm{out}}J_{V}^{h_{i}}+J_{V}^{+}\Big)\braout \fpgrav \ \sm\ketin
\end{align}
where $\hat{x}$ denotes the direction of the soft graviton parametrized by $(w_{1},\bar{w_{1}})$ on the conformal sphere.  $J_{V}^{+}$ represents the action of $
\qvhard$ on the soft graviton with energy $E_{p}$.

Similarly, the second term in \eqref{ward3al} can be evaluated to:
\begin{align}
\braout\qfsoft\sm\qvhard\ketin=\frac{i}{2\pi}\lim_{E_{p}\rightarrow 0}\int d^{2}{w_{1}}~D^{2}_{\bar{w_{1}}}f~\Big(\sum_{\mathrm{in}}J_{V}^{-h_{i}}\Big)E_{p}\braout a_{+}(E_{p}\hat{x}) \ \sm\ketin
\end{align}

Hence, the Ward identity \eqref{ward3al} simplifies to:
\begin{align}\nonumber
\braout\qfsoft \qvsoft \sm\ketin=&\\\nonumber-\frac{i}{2\pi}\lim_{E_{p}\rightarrow 0}\int d^{2}{w_{1}}& D^{2}_{\bar{w_{1}}}f  \Big(\sum_{\mathrm{out}}J_{V}^{h_{i}}-\sum_{\mathrm{in}}J_{V}^{-h_{i}}\Big)\Big[E_{p}\braout a_{+}(E_{p}\hat{x}) \sm\ketin\Big]\\\label{sl1a}
-\frac{i}{2\pi}&\lim_{E_{p}\rightarrow 0}\int d^{2}{w_{1}}~D^{2}_{\bar{w_{1}}}f~E_{p}\Big(J_{V}^{+}\Big)\Big[\braout a_{+}(E_{p}\hat{x}) \ \sm\ketin\Big]
\end{align}
Note that, the l.h.s. of \eqref{sl1a} can be written as:\footnote{More precise definition of l.h.s. is given in Appendix \ref{careful}.}
\begin{equation}\label{anupam_warda}
\begin{split}
\lim_{E_{p}\rightarrow 0 } \ \frac{1}{2\pi} E_{p} \ \lim_{E_{q}\rightarrow 0 } \ \frac{1}{4\pi i}(&1 + E_{q}\partial_{E_{q}}) \times 
\\&\int d^{2}{w_{1}} \ d^{2}{w_{2}} \ D^{2}_{\bar{w_{1}}}f \  \partial^{3}_{\bar{w_{2}}}V^{\bar{w_{2}}}\braout\fpgrav \ \vpgrav \ \sm\ketin
\end{split}
\end{equation}
It is important to note that the soft limits taken in the above equation do not follow any particular order in the energies of the soft gravitons. However as we show in the next section, the right hand side of the Ward identity is equivalent to the right hand side of one of the CDSTs .

\subsubsection{From Ward Identity to Soft Theorem}
Having derived the Ward identity \eqref{sl1a}, we now ask whether it can be interpreted as a soft theorem.
Motivated by the single soft graviton case, we make the following choices for function $f$ and vector field $V$:
\begin{align}\label{choice}
f(w_{1},\bar{w_{1}})&=s(w_{1},\bar{w_{1}};w_{p},\bar{w_{p}})\\
V^{\bar{w_{2}}}&=K^{+}_{(w_{q},\bar{w_{q}})}\nonumber
\end{align} 
where $s(w_{1},\bar{w_{1}};w_{p},\bar{w_{p}})$ and $K^{+}_{(w_{q},\bar{w_{q}})}$ follow the definitions in Section \ref{charges}. Using this, \eqref{anupam_warda} becomes:
\begin{align}
\lim_{E_{p}\rightarrow 0}E_{p}\lim_{E_{q}\rightarrow 0}(1+E_{q} \ \partial_{E_{q}})\braout\fpgrav \ \vpgrav \ \sm\ketin
\end{align}
where the unit vectors $\hat{x}$ and $\hat{y}$ denote the coordinates $(w_{p},\bar{w_{p}})$ and $(w_{q},\bar{w_{q}})$ on the conformal sphere.

Further, for the r.h.s. of \eqref{sl1a}, we have: 
\begin{equation}\label{soft1a}
\begin{split}
\lim_{E_{p}\rightarrow 0}\sum\limits_{i} S^{(1)}(q;k_{i})\Big[E_{p} \ \braout & a_{+}(E_{p}\hat{x}) \ \sm\ketin\Big]
\\
&+ \lim_{E_{p}\rightarrow 0}E_{p} \ S^{(1)}(q;p)\Big[\braout a_{+}(E_{p}\hat{x}) \ \sm\ketin\Big]
\end{split}
\end{equation}
In the above expression, notice that in both the subleading factors $S^{(1)}(q;k_{i})$ and $S^{(1)}(q;p)$,  the soft graviton with momentum $q$ is localized at $\hat{y}$ on the conformal sphere. However, the first one contains an angular momentum operator acting on the $i^{\mathrm{th}}$ hard particle and the latter contains an angular momentum operator acting on the soft graviton with momentum $p$.

Now, using the leading single soft theorem, the first term in \eqref{soft1a} can be written as:
\begin{align}
\sum\limits_{i}S^{(1)}(q;k_{i})\Bigg[\sum\limits_{j}E_{k_{j}} \hat{S}^{\0}(p;k_{j})\braout\sm\ketin\Bigg]
\end{align}

For the second term in \eqref{soft1a}, we use the expansion of the $(n+1)$ particle amplitude \eqref{cons1} and we get a factorization of the form:
\begin{align}\label{sl2}
\braout a_{+}(E_{p}\hat{x}) \ \sm\ketin=\Bigg[\sum\limits_{i}\frac{E_{k_{i}}}{E_{p}} \ \hat{S}^{\0}(p;k_{i})+\sum\limits_{i}S^{(1)}(p;k_{i})\Bigg]\sa+\mathcal{O}(E_{p})
\end{align}

The second term of \eqref{sl2} is at a higher order in soft graviton energy, and so does not contribute to \eqref{soft1a}. Thus, \eqref{soft1a} finally becomes:
\begin{equation} \label{soft2a}
\begin{split}
\sum\limits_{i}S^{(1)}(q;k_{i})\Bigg[\sum\limits_{j}E_{k_{j}} \hat{S}^{\0}(p;k_{j})&\braout\sm\ketin\Bigg]\\+\lim_{E_{p}\rightarrow 0}~E_{p}& \ S^{(1)}(q;p)\Bigg[\sum\limits_{j}\frac{E_{k_{j}}}{E_{p}} \ \hat{S}^{\0}(p;k_{j})\Bigg]\braout\sm\ketin
\end{split}
\end{equation}

Lastly, since $S^{(1)}(q;k_{i})$ is a linear differential operator and $S^{(1)}(q;p)$ acts only on the soft coordinates, we can further simplify \eqref{soft2a} as:
\begin{equation}\label{softtheormalternatea}
\begin{split}
\Bigg[\sum\limits_{i,j}E_{k_{i}} \ \hat{S}^{\0}(p;k_{i}) \ S^{(1)}(q;k_{j})+\sum\limits_{i}& S^{(1)}(q;k_{i})\Big(E_{k_{i}} \ \hat{S}^{\0}(p;k_{i})\Big) 
\\ 
+ \lim_{E_{p}\rightarrow 0}~E_{p}& \ S^{(1)}(q;p)\Big(\sum\limits_{j}\frac{E_{k_{j}}}{E_{p}} \ \hat{S}^{\0}(p;k_{j}) \Big)\Bigg]\sa
\end{split}
\end{equation}

Finally, putting this all together, we get a subleading double soft theorem:
\begin{equation}\label{slst1a}
\begin{split}
\lim_{E_{p}\rightarrow 0 }E_{p} &\lim_{E_{q}\rightarrow 0}(1+E_{q} \ \partial_ {E_{q}}) \ \mathcal{A}_{n+2}(q,p,\{k_{m}\})\\
=&\Bigg[S^{\0}(p;\{k_{i}\}) \ S^{\1}(q;\{k_{j}\}) + \mathcal{M}_{1}(q;p;\{k_{i}\}) + \mathcal{M}_{2}(q;p;\{k_{i}\})\Bigg]\mathcal{A}_{n}(\{k_{m}\})
\end{split}
\end{equation}
where, $\mathcal{M}_{1}(q;p;\{k_{i}\})$ and $\mathcal{M}_{2}(q;p;\{k_{i}\})$ are the same contact terms obtained in subleading CDST \eqref{contact2}, whose expressions can be read off from \eqref{ttt-1}, \eqref{ttt-2} respectively. This is the same subleading consecutive double soft theorm \eqref{contact2}, that we studied in the Section \ref{Consecutive}. Note however that, in \eqref{anupam_warda} there is no particular ordering in the limits of the soft graviton energy obtained from the successive action of the soft charges. Hence, the l.h.s. of the double soft theorem \eqref{slst1a} contains independent limits as opposed to \eqref{contact2}, where the limits have definite ordering.  Although we believe this point needs to be better understood,  what we have shown here is that the Ward identity of superrotation charges in a supertranslated vacuum leads to a particular CDST.
It is also important to emphasise that there is a definite time ordering in $\big[Q_{f},[Q_{V},\mathcal{S}] \big]\ =\ 0$. This is clear from  the derivation of the Ward identity $\bra{\mathrm{out}} \big[  Q_{f},  [Q_{V},\mathcal{S}] \big]\ket{\mathrm{in}}=0$, which is presented in Appendix \ref{derieveconsecutive}.

\section{ Relating the Standard CDST to a Ward Identity}\label{scdst}

As we saw above, the Ward identity $ [Q_{f}, [Q_{V}, \mathcal{S}]]=0$, gave rise to a double soft theorem whose r.h.s. matched with the consecutive soft theorem, where we considered the subleading limit of the graviton which was taken soft first. This is in contrast to the more standard consecutive soft limit where we consider the leading soft limit of the graviton which is taken soft first  and subleading soft limit of the graviton which is taken soft second. We will argue how this CDST could potentially arise out of the Ward identity:
\begin{align}\label{sub-ward}
\bra{\mathrm{out}}\big[Q_{V} , [Q_{f}, \mathcal{S}]\big]\ket{\mathrm{in}}=0
\end{align}
Expressing the charges in \eqref{sub-ward} as the sum of hard and soft charges, we get:
\begin{equation}\label{sub-ward2}
\begin{split}
\bra{\mathrm{out}} \big[\qvhard,& [ \qfhard, \mathcal{S}]\big] \ket{\mathrm{in}} + \bra{\mathrm{out}} \big[\qvsoft, [\qfhard, \mathcal{S}]\big] \ket{\mathrm{in}} 
\\
&+ \bra{\mathrm{out}} \big[\qvhard, [\qfsoft, \mathcal{S}] \big] \ket{\mathrm{in}} + \bra{\mathrm{out}} \big[\qvsoft,[ \qfsoft, \mathcal{S}] \big] \ket{\mathrm{in}}=0
\end{split}
\end{equation}
Using the Ward identity for  supertranslation, namely $[\qfsoft , \mathcal{S}] = - [\qfhard, \mathcal{S}]$, the first and the third terms cancel each other.  Once again, this leads us to the following supertranslation Ward identity evaluated in states defined with respect to ``super--rotated vacuum".
\begin{equation}
\begin{split}
\bra{\mathrm{out}} Q_{V}^{\textrm{soft}}\ [Q_{f},\mathcal{S}] \ket{\mathrm{in}} &= 0
\\
\bra{\textrm{out}, V} [Q_{f}, \mathcal{S}] \ket{\mathrm{in}} &= 0
\end{split}
\end{equation}
where by $\ket{\textrm{out},V}$ we mean a finite energy scattering state defined with respect to a vacuum which contains a subleading soft graviton mode.\footnote{It was shown in \cite{alok:new} how $Q_{V}^{\textrm{soft}}$ maps the vacuum to a different vacuum.} However, as we explain in appendix B,  unlike the action of $Q_{f}^{\textrm{soft}}$, the action of $Q_{V}^{\textrm{soft}}$ is not well understood thus far.\footnote{We are indebted to Prahar Mitra for emphasizing this point.} Consequently, the proposed Ward identity remains rather formal at this point. We will still proceed further and show that this proposed Ward identity, if well defined is equivalent to the standard CDST. 
We can rewrite the Ward identity as
\begin{equation} \label{master}
\begin{split}
\braout\qvsoft \ &\qfsoft \ \sm\ketin
\\
&=-\braout[\qvsoft,[\qfhard,\sm]]\ketin
\\
&=\braout\qvsoft\sm\qfhard-\qfhard\qvsoft\sm\ketin+\braout[\qfhard,\qvsoft]\sm\ketin
\end{split}
\end{equation}

We evaluate the two terms in the r.h.s. of \eqref{master} one by one. The first term can be written as:
\begin{equation}\label{hardhard}
\begin{split}
\braout\qvsoft \ \sm \ \qfhard-&\qfhard \ \qvsoft \ \sm\ketin
\\
&=-\braout[\qfhard,\qvsoft \ \sm]\ketin =-\braout \big[\qfhard,[\qvsoft,\sm] \big]\ketin
\\
&=\braout \big[\qfhard,[\qvhard,\sm] \big]\ketin
\end{split}
\end{equation}
Then, using the action of $\qfhard$ and $\qvhard$ on the external states, we can write the r.h.s. of \eqref{hardhard} as:
\begin{align}\label{part1}
~~~\braout\big[\qfhard,[\qvhard,\sm]\big]\ketin=&\\i\Bigg[\sum_{\mathrm{out}} f(\hat{k_{i}})E_{i}-\sum_{\mathrm{in}}& f(\hat{k_{i}})E_{i}\Bigg]\Bigg[\sum_{\mathrm{out}} J^{h_{i}}_{V_{i}}-\sum_{\mathrm{in}}J^{-h_{i}}_{V_{i}}\Bigg]\bra{\mathrm{out}}\mathcal{S}\ket{\mathrm{in}}\nonumber
\end{align}

To evaluate the second term in \eqref{master}, note that for a single particle state $\ket{k}$,
\begin{equation}\label{cvsfh}
\begin{split}
\bra{k} [\qfhard,& \ \qvsoft] 
\\
&= -\frac{1}{4\pi i} \lim_{E_{p} \rightarrow 0} (1 + E_{p} \ \partial_{E_{p}}) \int d^{2}w_{2} \  \partial^{3}_{\bar{w_{2}}} V^{\bar{w_{2}}} \  E_{p}  \ f(w_{2},\bar{w_{2}}) \bra{k} a_{+}(E_{p}, w_{2}, \bar{w_{2}}) 
\\
&= -\frac{1}{4\pi i} \lim_{E_{p}\rightarrow 0} \int d^{2}w_{2} \   \partial^{3}_{\bar{w_{2}}} V^{\bar{w_{2}}} \ E_{p} \ f(w_{2},\bar{w_{2}}) \bra{k} a_{+}(E_{p}, w_{2}, \bar{w_{2}})
\end{split}
\end{equation}
Where,  in going from the first line to the second, we have used the fact that $a_{+}(E_{p}, w_{2}, \bar{w_{2}})\sim\frac{1}{E_{p}}$.\footnote{This can be seen by writing the mode functions of News tensor ($N^{\omega}_{zz}$), in terms of graviton annihilation operators as in \cite{strom}.} Therefore,
\begin{align}
-\frac{1}{4\pi i}\lim_{E_{p}\rightarrow 0} E_{p} \ \partial_{E_{p}}\int d^{2}w_{2} \  \partial^{3}_{\bar{w_{2}}} V^{\bar{w_{2}}} \  E_{p}  \ f(w_{2},\bar{w_{2}}) \bra{k} a_{+}(E_{p}, w_{2}, \bar{w_{2}})=0
\end{align}
Using the above expression \eqref{cvsfh}, we can evaluate the second term of (\ref{master}) as:
\begin{align}\label{new-master2}
\bra{\mathrm{out}} [\qfhard, \qvsoft] \ \mathcal{S} \ket{\mathrm{in}}= -\frac{1}{4\pi i} \lim_{E_{p} \rightarrow 0} \int d^{2}w_{2} \ & \partial^{3}_{\bar{w_{2}}} V^{\bar{w_{2}}} \ E_{p} \\& \times f(w_{2},\bar{w_{2}}) \bra{\mathrm{out}} a_{+}(E_{p}, w_{2}, \bar{w_{2}}) \ \mathcal{S} \ket{\mathrm{in}}\nonumber
\end{align}
Lastly, using the single soft graviton theorem (with energy $E_{p}$), \eqref{new-master2} simplifies to:
\begin{equation}\label{new-master3}
\begin{split}
\bra{\mathrm{out}}[\qfhard, \qvsoft] \ &\mathcal{S} \ket{\mathrm{in}}
\\
& =-\frac{1}{4\pi i} \sum\limits_{i} \int d^{2}w_{2} \  \partial^{3}_{\bar{w_{2}}}V^{\bar{w_{2}}}  \ f(w_{2},\bar{w_{2}}) \ E_{k_{i}} \ \hat{S}^{(0)}(p;k_{i}) \sa
\end{split}
\end{equation}
Finally, substituting (\ref{part1}) and (\ref{new-master3}) in (\ref{master}), we arrive at the Ward identity:
\begin{equation}\label{slcdls}
\begin{split}
\bra{\mathrm{out}} &\qvsoft \  \qfsoft \  \mathcal{S} \ket{\mathrm{in}} 
\\
& = i\Bigg[\sum_{\mathrm{out}} f(\hat{k_{i}}) E_{i} - \sum_{\mathrm{in}} f(\hat{k_{i}})E_{i}\Bigg] \Bigg[\sum_{\mathrm{out}} J^{h_{i}}_{V_{i}} - \sum_{\mathrm{in}} J^{-h_{i}}_{V_{i}}\Bigg] \bra{\mathrm{out}} \mathcal{S} \ket{\mathrm{in}} 
\\
& \quad -\frac{1}{4\pi i} \sum\limits_{\mathrm{hard}} \int d^{2}w_{2} \  \partial^{3}_{\bar{w_{2}}}V^{\bar{w_{2}}}  \ f(w_{2},\bar{w_{2}}) \ E_{k_{i}} \ S^{(0)}(w_{2}, \bar{w_{2}};k_{i}) \sa
\end{split}
\end{equation}
where the l.h.s. can be expressed as:
\begin{align}\nonumber\label{fdswi}
\frac{1}{4\pi i}\lim_{E_{p} \rightarrow 0} \big(1+E_{p} \ \partial_{E_{p}}\big)  \ \frac{1}{2\pi}\lim_{E_{q}\rightarrow 0}  E_{q} \int d^{2}w_{1} \ &d^{2}w_{2} ~ D_{\bar{w}_{1}}^{2}f(w_{1},\bar{w_{1}})~ \partial_{\bar{w}_{2}}^{3}V^{\bar{w_{2}}}&\\
\times\bra{\mathrm{out}} & a_{+}(E_{q},w_{1},\bar{w}_{1}) \  a_{+}(E_{p},w_{2},\bar{w}_{2}) \ \mathcal{S}\ket{\mathrm{in}}
\end{align}

In order to proceed from the Ward identity \eqref{slcdls} to a soft theorem we make the following choices for $f$ and $V$:
\begin{align}\label{key}
f(w_{1},\bar{w_{1}}) = s(w_{1}, \bar{w_{1}};w_{q}, \bar{w_{q}}) \  , \ \ V^{\bar{w_{2}}}= K^{+}_{(w_{p},\bar{w_{p}})}
\end{align}
Substituting these in \eqref{slcdls}, we formally get the subleading CDST for positive helicity gravitons as:
\begin{align}\label{consecutive1}\nonumber
\lim_{E_{p} \rightarrow 0}& \big(1+ E_{p} \ \partial_{E_{p}} \big) \lim_{E_{q} \rightarrow 0} E_{q}  \bra{\mathrm{out}} a_{+}(E_{q}\hat{y}) \ a_{+}(E_{p}\hat{x}) \ \mathcal{S}\ket{\mathrm{in}} \  ~~~~~~~~~~~~~~~~~~~~~~ 
\\
&= \bigg[S^{(0)}(q;\{k_{i}\}) \ S^{(1)}(p;\{k_{j}\})
+ \hat{S}^{(0)} (q;p) \ S^{(0)}(p;\{k_{i}\})\bigg]\sa
\end{align}
Again, $\hat{x}$ and $\hat{y}$ denote the points $(w_{p},\bar{w_{p}})$, $(w_{q},\bar{w_{q}})$ on the conformal sphere. This is the same consecutive double soft theorem \eqref{proj1} discussed in Section \ref{Consecutive}.\\

However, as discussed in Appendix \ref{careful}, there are some important subtleties in the definition of soft operators, especially the soft super-rotation charge  $Q^{\mathrm{soft}}_{V}$. Due to this, in the evaluation of the Ward identity $\bra{\mathrm{out}}[Q_{V} , [Q_{f}, \mathcal{S}]]\ket{\mathrm{in}}=0$, the steps which involve the operation of charge $Q^{\mathrm{soft}}_{V}$ first on the ``out" state before the other charge are not mathematically rigorous.  However, we present this calculation here, in the hope that this might give some hint to the structure of a more mathematically sound proof of this soft theorem as well as a more rigorous understanding of the operation of the soft superrotation charge. 
\section{Discussion and Conclusion}
It has now been well established in the literature that the supertranslation soft charge $Q^{\textrm{soft}}_{f}$ shifts the Fock Vacuum to a  vacuum parametrized by a soft graviton. If we consider Ward identities associated to superrotation charges $Q_{V}$ in this supertranslated vacuum, we are led to one of the two consecutive subleading double soft graviton theorems. In fact, as was argued in \cite{alok:new}, the space of vacua of (perturbative) Quantum Gravity are parametrized by leading as well as subleading soft gravitons. Although we do not have a precise definition of a vacuum which is labelled by a subleading soft graviton, assuming such a definition exists, we can ask what the Ward identity of the supertranslation charge is in such a state. The answer appears to be related to the other consecutive double soft theorem at the subleading level.

Many questions remain open. A precise formulation of these Ward identities will require a careful definition of $Q_{V}^{\textrm{soft}}$  which is lacking thus far. It is also not entirely clear why Ward identity associated to $Q_{V}$ ``in" states perturbed around the supertranslated vacuum leads to a specific CDST.

It will also be interesting to extend the analysis to the case where the finite energy scattering states are massive. This will require a detailed understanding of the BMS algebra at time--like infinity. Finally, the problem of relating the subleading simultaneous double soft theorem to Ward identities associated to Asymptotic symmetries remain completely open. Based on our analysis above, we expect that this will require a detailed analysis of the moduli space of the vacuua (parametrized by leading and subleading soft gravitons) which is complicated by the non--Abelian nature of the BMS symmetries. 

\section*{Acknowledgement:} The authors are thankful to Alok Laddha for defining the problem,  crucial and insightful discussions at various stages, help with the calculations and writing the manuscript. We are also thankful to Prahar Mitra for key conceptual discussions regarding the (to date unresolved issue of defining the) domain of superrotation charge and difficulties in recovering both the consecutive soft limits from Ward identities. We would like to thank Ashoke Sen for discussions regarding multiple soft limits. AAH would like to thank Renjan John and Arnab Priya Saha for helpful discussions and for comments on the manuscript. Much of this work was completed when KR was a Long Term Visiting Student at the International Centre for Theoretical Sciences, Bangalore --  he would like to thank them for their gracious support and hospitality. KR would also like to thank the Chennai Mathematical Institute for granting him permission to visit ICTS for this period.
\bigskip
\appendix
\section{Ward Identities from the Avery--Schwab Method}\label{derieveconsecutive}

In this appendix we derive the asymptotic Ward identity $\bra{\textrm{out}}\big[Q_{f}, [Q_{V},\mathcal{S}]\big]\ket{\mathrm{in}} = 0$, based on a method that was proposed in \cite{avery}. The basic idea is to  use Noether's second theorem and path integral techniques to derive Ward identities for asymptotic symmetries.

As shown in \cite{avery}, given a asymptotic symmetry or large gauge transformation with a gauge parameter $\lambda$, at the level of correlation functions one obtains the following Ward identity.
\begin{equation}\label{as1}
\begin{split}
-i \bra{0} \delta_{\lambda} T\Big(\Phi (x_{1}) \dots \Phi(x_{n}) \Big) \ket{0} = \bra{0} T\Big( \big(Q_{{\cal I}^{+}}[\lambda]\ -\ Q_{{\cal I}^{-}}[\lambda]\big) \Phi (x_{1}) \dots \Phi(x_{n}) \Big)\ket{0}
\end{split}
\end{equation}

Here we use a generic label $\Phi$ to label the quantum field associated to scattering particles.  $Q_{{\cal I}^{\pm}}[\lambda]$ are the asymptotic charges associated to large gauge transformations $\lambda$ at future and past null infinity respectively. 

Before deriving the identity associated to the insertion of two charge operators, we first revisit the supertranslation Ward identity  $\bra{\textrm{out}} [Q_{f}, \mathcal{S}]\ket{\textrm{in}} = 0$. Let $\Phi$ be any massless field that interacts with gravity and $\delta_{\lambda} = \delta_{f}$ be the generator of supertranslation on the fields.
 
We begin by noting  that through LSZ reduction we have the following\footnote{These arguments are formal because they are tied to the fact that the usual Dyson $\mathcal{S}$--matrix with massless particles is only formally defined. However, as we are only analyzing symmetries of the tree--level $\mathcal{S}$--matrix, we will not worry about the issue of infra--red divergence.}
\begin{equation}
\begin{split}
\prod_{i=1}^{m}p_{i}^{2}\ \int d^{4}x_{i}\ e^{-ip_{i}\cdot x_{i}} &\prod_{j=m+1}^{n} p_{j}^{2}\int d^{4}x_{j}\ e^{ip_{j}\cdot x_{j}}
\bra{0} \delta_{f}T\left(\Phi(x_{1})\dots\Phi(x_{n})\right) \ket{0} 
\\
&=-i\bra{p_{1},\dots,p_{m}} Q^{\textrm{hard} +}_{f}\ \mathcal{S} - \mathcal{S}\ Q^{\textrm{hard} -}_{f}\ket{p_{m+1},\dots,p_{n}}
\end{split}
\end{equation}
We can schematically represent this step as,
\begin{equation}\label{lsz1}
\begin{array}{lll}
\bra{0} \delta_{\lambda} T\left(\Phi (x_{1})\ \dots\ \Phi(x_{n})\right)\ket{0}  \xrightarrow[\mathrm{LSZ}] \ \bra{p_{1},\dots,p_{m}}  [Q_{f}^{\textrm{hard}},\mathcal{S}]\ \ket{p_{m+1},\dots,p_{n}}
\end{array}
\end{equation}
where we have used the fact that 
\begin{equation}\label{av-sch-0}
\delta_{f}\Phi(p) = -i \ [Q_{f},\Phi(p)]
\end{equation}

On the other hand, once again via LSZ and the fact that 
\begin{equation}
\begin{split}
Q^{\textrm{hard}}_{f} \ket{0}&= 0
\\
Q^{\textrm{soft}}_{f} \ket{0} &= 0
\\
\bra{0} Q_{f}^{\textrm{soft}} &\neq 0
\end{split}
\end{equation}
we see that
\begin{equation}\label{lsz2}
\begin{split}
\bra{0} T\Big( \big(Q_{{\cal I}^{+}}[\lambda]-Q_{{\cal I}^{-}}[\lambda]\big)\Phi (x_{1})\dots\Phi(x_{n})\Big) \ket{0}  \xrightarrow[\textrm{LSZ}]  \ 
\bra{ p_{1},\dots,p_{m}} [Q_{f}^{\textrm{soft}},\mathcal{S}] \ket{p_{m+1},\dots,p_{n}} 
\end{split}
\end{equation}

Substituting eqns. (\ref{lsz1},\ref{lsz2}) in eq.(\ref{as1}) we recover the super-translation Ward identity,
\begin{equation}
\bra{\textrm{out}} [Q_{f},\mathcal{S}] \ket{\textrm{in}} =\ 0
\end{equation}

We note that an identical derivation for Ward identity associated to large $U(1)$ gauge transformations was already given in \cite{strominger-ym1}.\\

We will now derive the Ward identities $\big[Q_{f}, [Q_{V}, \mathcal{S}]\big] = 0$ using this method. That is, we begin with the Ward identity where the superrotation $\delta_{V}$ is applied after the supertranslation $\delta_{f}$. The starting point for the derivation is (45) in \cite{avery}, which in the present context can be written as 
\begin{equation}\label{av-sc-1}
\begin{split}
-\bra{0} T \Big( \big(Q_{{\cal I}^{+}}[f] \ - \ Q_{{\cal I}^{-}}[f] \big) \big(Q_{{\cal I}^{+}}[V]\ -\ Q_{{\cal I}^{-}}&[V]\big) \Phi(x_{1})\dots\Phi(x_{n}) \Big)\ket{0} 
\\
=& \bra{0} \delta_{f} \ \delta_{V} T\Big(\Phi(x_{1})\dots\Phi(x_{n})\Big) \ket{0}
\end{split}
\end{equation}

With our prescription that the soft charges annihilate the ``in" vacuum, the l.h.s. of (\ref{av-sc-1}) reduces to
\begin{equation}\label{av-sc-2}
\begin{split}
-\bra{0} T \Big( \big(Q_{{\cal I}^{+}}&[f] \ - \ Q_{{\cal I}^{-}}[f] \big) \big(Q_{{\cal I}^{+}}[V]\ -\ Q_{{\cal I}^{-}}[V]\big) \Phi(x_{1})\dots\Phi(x_{n}) \Big)\ket{0} 
\\
&= -\bra{0} Q^{\textrm{soft}}_{{\cal I}^{+}}[f]\ \Big(Q^{\textrm{soft}}_{{\cal I}^{+}}[V]\ +\ Q^{\textrm{hard}}_{{\cal I}^{+}}[V]\Big) T\big( \Phi(x_{1})\dots\Phi(x_{n}) \big) \ket{0}
\end{split}
\end{equation}
On the other hand, using (\ref{av-sch-0}), it is easy to see that the r.h.s. of (\ref{av-sc-1}) is given by
\begin{equation}\label{av-sc-3}
\begin{split}
\bra{0} \delta_{f} \ \delta_{V}& \ T\left(\Phi(x_{1})\dots\Phi(x_{n})\right)\ \ket{0}
\\
&=-\bra{0}  \sum_{i,j} T\left(\Phi(x_{1})\dots[Q_{f},\Phi(x_{i})]\dots[Q_{V},\Phi(x_{j})]\dots\Phi(x_{n})\right)\ket{0}
\\
& \qquad \qquad \qquad  \qquad \qquad \qquad  \qquad \xrightarrow[\textrm{LSZ}] \ - \bra{\textrm{out}} \big[ Q_{f}^{\textrm{hard}},[Q_{V}^{\textrm{hard}},\ \mathcal{S}] \big] \ket{\textrm{in}}
\end{split}
\end{equation}

Thus the path integral identity and the LSZ formula lead to (equating the r.h.s. of (\ref{av-sc-2}) with r.h.s. of (\ref{av-sc-3})),
\begin{equation}
\begin{array}{lll}
\bra{\textrm{out}} Q^{\textrm{soft}}_{f} \ &Q^{\textrm{soft}}_{V} \ \mathcal{S} \ket{\textrm{in}} 
\\
&= - \bra{\textrm{out}} Q^{\textrm{soft}}_{f} \ Q_{V}^{\textrm{hard}} \ \mathcal{S}\ \ket{\textrm{in}} \ + \ \bra{\textrm{out}} \big[ Q_{f}^{\textrm{hard}},[Q_{V}^{\textrm{hard}},\ \mathcal{S}] \big] \ket{\textrm{in}}
\end{array}
\end{equation}

A straightforward manipulation shows that above equation is equivalent to
\begin{equation}
\bra{\textrm{out}} \big[Q_{f},[Q_{V}, \mathcal{S}] \big] \ket{\textrm{in}} =\ 0
\end{equation}
This is one of the Ward identities used in the main text of the paper. The remaining identites can be derived similarly.

\section{Subtleties Associated to the Domain of Soft Operators}\label{careful}

We will now comment on the assumption that was implicitly used in previous section, and  which has been used frequently in relating single soft theorems to BMS Ward identities.\footnote{The authors would like to thank Abhay Ashtekar and Miguel Campiglia for explaining this subtlety to us in detail in the context of supertranslations, and Prahar Mitra for patiently explaining to us why this subtlety cannot be avoided when we look at Ward identities associated to double soft theorems \cite{abhay-miguel},\cite{prahar-ym}.}

From the expressions of the supertranslation and superrotation soft charges, we can see that these are 
singular limits of single graviton  annihilation operators. 
\begin{align}\nonumber
Q_{f}^{\textrm{soft}}\ &\sim\ \lim_{E\rightarrow 0}E ~a_{+}(E,w,\bar{w})\\
Q_{V}^{\textrm{soft}}\ &\sim\ \lim_{E\rightarrow 0}~(1 + E\partial_{E}) a_{+}(E,w,\bar{w})
\end{align}

For simplicity we have just considered the expression of the soft charges for positive helicity graviton creation operators only. In the case of Ward identities associated to the single soft theorems, it has been implicitly assumed that  the super-translation soft charge can be defined as (apart from the extra factors),
\begin{align}
\bra{\mathrm{out}}\lim_{E \rightarrow 0}E \ a_{+}(E,w,\bar{w}) \ \mathcal{S}\ket{\mathrm{in}}=\lim_{E \rightarrow 0}E\bra{\mathrm{out}}a_{+}(E,w,\bar{w}) \ \mathcal{S}\ket{\mathrm{in}}
\end{align}
A similar assumption is also made for the superrotation soft charge $Q_{V}^{\textrm{soft}}$.\\

However, this does not take into account the fact that the supertranslation soft charge shifts the vacuum. This subtlety is now well understood for supertranslations. It was shown in \cite{sever,strom-fadeev,akhoury1,akhoury2} that the action of the supertranslation soft charge maps a standard Fock vaccuum to a supertranslated state which can be thought of as being labelled by a single soft graviton. With this is in mind the precise definition of $\langle\textrm{out}\vert Q_{f}^{\textrm{soft}} \ Q_{V}^{\textrm{soft}}\ \mathcal{S}\ \vert\textrm{in}\rangle$ would be
\begin{align}
\braout Q_{f}^{\textrm{soft}} \ Q_{V}^{\textrm{soft}}\ \mathcal{S} \ketin : \approx \int d^{2}w\ D^{3}_{\bar{w}}V^{\bar{w}} \ \bra{\textrm{out},f}\lim_{E\rightarrow 0}\ (1 + E\partial_{E})\ a_{+}(E,w,\bar{w}) \ \mathcal{S} \ketin 
\end{align}
where $\langle \textrm{out},f\vert$ is the ``out'' state defined over the shifted vaccuum parametrized by $f$, generated by the action of supertranslation charge ($\qfsoft$) on the Fock vaccuum. 
\\

 In  going from \eqref{sl1a} to \eqref{anupam_warda} we have made the same assumption for defining $Q_{V}^{\textrm{soft}}$ on the shifted vacuum as has been made in the literature for defining it on the Fock vacuum, namely:

\begin{align}
\bra{\textrm{out},f} \lim_{E\rightarrow 0}\ (1 + E\partial_{E})\ a_{+}(E,w,\bar{w})\ :=\ 
\lim_{E\rightarrow 0}\ (1 + E\partial_{E})\ \bra{\textrm{out},f} a_{+}(E,w,\bar{w})
\end{align}

 However for reasons which can be traced back to the classical theory, it is still not clear what the precise definition of $Q_{V}^{\textrm{soft}}$ is. That is, just as a rigorous definition of $Q_{f}^{\textrm{soft}}$ being defined as an operator which maps the ordinary Fock vacuum to a super--translated state \cite{strom-fadeev,akhoury1}, no corresponding definition is available for $Q_{V}^{\textrm{soft}}$ as yet. Consquently, operator insertions like $\langle\textrm{out}\vert Q_{V}^{\textrm{soft}} \ Q_{f}^{\textrm{soft}} \ \mathcal{S}\ \vert\textrm{in}\rangle$ are not mathematically well--defined, and we do not know how to make sense of them.

\end{document}